\documentclass[letterpaper,twocolumn,10pt]{IEEEtran}

\usepackage{tikz}
\usepackage{amsmath}
\usepackage{graphicx}
\usepackage{enumitem}
\setlist[itemize]{leftmargin=*}
\usepackage{xfrac}
\usepackage[affil-it]{authblk}
\usepackage{filecontents}
\usepackage{xcolor}
\usepackage{etoolbox}
\usepackage{xspace}
\usepackage{caption}
\usepackage{subcaption}
\usepackage{booktabs}
\usepackage{tabularx}
\usepackage{multirow}

\usepackage{balance}

\newbool{showcomments}
\booltrue{showcomments}

\newcommand{\mycomment}[3]{
    \ifbool{showcomments}{
        {\itshape\small\color{#2}[#1:#3]}
    }{}
}

\begin{document}

\date{}

\title{\Large \bf The Multipath Reliable Connection (MRC) Transport}
\makeatletter
\renewcommand\AB@affilsepx{, \protect\Affilfont} 
\makeatother
\renewcommand\Authsep{, }
\renewcommand\Authand{, }
\renewcommand\Authands{, }

\author[1]{Rip Sohan}
\author[2]{Eric Spada}
\author[2]{Eric Davis}
\author[5]{Mark Handley}
\author[4]{Idan Burstein}
\author[6]{Tony Hurson}
\author[3]{Jithin Jose}
\author[6]{Vivek Kashyap}
\author[1]{Rong Pan}
\author[4]{Sayantan Sur}

\author[3]{Sreevatsa Anantharamu}
\author[4]{Aviv Barnea}
\author[5]{Adrian Caulfield}
\author[4]{Elazar Cohen}
\author[4]{Elliot Edmunds}
\author[4]{Yamin Friedman}
\author[3]{Mahdieh Ghazi}
\author[1]{Murali Garimella}
\author[3]{Torsten Hoefler}
\author[1]{Vipin Jain}
\author[3]{Abdul Kabbani}
\author[4]{Noam Katz}
\author[1]{Yanfang Le}
\author[4]{Charlie Mbariky}
\author[2]{Guglielmo Morandin}
\author[4]{Masoud Moshref}
\author[1]{Shane O'Neil}
\author[5]{Michael Papamichael}
\author[4]{Jonas Pfefferle}
\author[1]{Siva Santosh Pyla}
\author[2]{Costin Raiciu}
\author[1]{David Riddoch}
\author[2]{Karen Schramm}
\author[4]{Yuval Shpigelman}
\author[4]{Shahaf Shuler}
\author[4]{Shy Shyman}
\author[1]{Raghava Sivaramu}
\author[5]{Amin Tootoonchian}
\author[3]{Yang Wang}

\affil[1]{Advanced Micro Devices Inc}
\affil[2]{Broadcom Inc}
\affil[3]{Microsoft Corp}
\affil[4]{NVIDIA Corp}
\affil[5]{OpenAI OpCo LLC}
\affil[6]{Intel Corp}

\maketitle

\begin{abstract}
MRC is an open, production-grade transport designed for large-scale
AI/ML training over best-effort Ethernet. It extends RoCEv2 with
explicit, composable primitives for per-packet multipath and
sender-based congestion control, decouples packet delivery from
semantic processing, adds multiple new capabilities for accelerated
packet-loss recovery and adds resilience against port and path
failures.  This paper presents MRC and details its core capabilities
and mechanisms.
\end{abstract}

\section{Introduction \& Motivation}

The network has emerged as a key bottleneck in large-scale AI/ML
training~\cite{meta-roce, dnntraining-overview, llama3}. As the
compute requirements and training cluster scales of frontier models
continue to grow, the network must sustain collective communication
across thousands of accelerators with low latency, high goodput, and
tolerance for failures. These requirements have driven significant
investment in high-performance interconnects, and industry has
responded with proprietary transport solutions~\cite{falcon,
  lu2025alibaba} as well as emerging clean-slate open
specifications~\cite{uet}. Proprietary transports, however, are
difficult to evolve across vendor boundaries, and clean-slate open
specifications require new hardware, new software stacks, and years of
interoperability maturation before they can be deployed at scale.

The dominant open standard in production AI/ML clusters today is
RoCEv2 Reliable Connection (RC) transport~\cite{ibta18}. RC provides
ordered, point-to-point delivery semantics and is well-understood by
hardware designers and software developers alike.  However, RC is
fundamentally insufficient for large-scale training: First, RC is
single-path with go-back-N retransmission, second, the RC congestion
control model (typically PFC-based lossless Ethernet or
DCQCN~\cite{dhen}) does not scale to large clusters~\cite{meta-roce}
and third, RC is highly vulnerable to link and fabric
failures~\cite{IRN}.

This paper presents the \emph{Multipath Reliable Connection} (MRC)
transport~\cite{mrc-spec}, an open specification developed under the
Open Compute Project (OCP). MRC takes a different approach from both
proprietary transports and clean-slate designs: it extends RC
directly, preserving its well-understood semantic and software model
while adding a coherent set of orthogonal primitives that address the
three gaps above. The result is a transport that can be adopted
incrementally, spanning design points from fixed-peer, single-tier
accelerator-integrated implementations to arbitrary-peer multi-tier
NIC deployments.

We make three contributions. First, we define an explicit
per-connection multipath and multiplane framework that enables
controlled packet spraying. Second, we specify reliability and
congestion control primitives suitable for best-effort Ethernet,
including selective acknowledgments, reasoned negative
acknowledgments, and wire-visible host backpressure signals. Third, we
introduce endpoint-scoped operations that provide scalable,
datapath-usable reachability and liveness signaling operations,
enabling fast failover without waiting for control-plane convergence.

MRC is already in production. A companion paper presents a full
empirical evaluation including goodput, latency, and failure-recovery
results~\cite{mrc-nsdi}. This paper describes the transport itself:
Section~\ref{sec:design} describes MRC's design and key mechanisms,
Section~\ref{sec:wire} summarizes the wire-protocol,
Section~\ref{sec:api} describes the controller and application API and
Section~\ref{sec:conclusion} concludes.

\begin{table*}[t]
\centering
\caption{MRC Transport Primitives: Mandatory (M) / Optional (O)\vspace{-0.07in}}
\label{tab:mechanisms}
\renewcommand{\arraystretch}{1.1}
\small
\begin{tabularx}{\textwidth}{l X c}
\toprule
\textbf{Feature} & \textbf{Description} & \textbf{M/O} \\
\midrule

\multicolumn{3}{l}{\bf \textit{Multipath \& Multiplane}} \\[1pt]
Per-connection packet spraying        & Rotates EVs on a per-packet
basis; algorithm is implementation-specific         & M \\
Source routing                        & Encodes path vectors using SRv6 uSIDs or MRC Structured EVs         & O \\
EV profiles                           & Controller-managed EV configuration shared across multiple QPs      & M \\
EV states \& denylisting              & Tracks path health; supports EV-scope disablement via datapath or central controller        & O \\
Multiplane operation                 & Enables multiport packet spraying isolated within a single QP                  & O \\[3pt]

\multicolumn{3}{l}{\bf \textit{Inflight Bounds}} \\[2pt]
Maximum PSN Range                    & Responder-advertised window bounding outstanding Packet Sequence Numbers   & M \\
Dynamic MPR                          & Runtime updates to connection MPR via responder SACK             & O \\
Maximum WriteIMM inflight            & Responder-advertised limit on concurrent inflight Write-Immediate operations    & M \\[3pt]

\multicolumn{3}{l}{\bf \textit{Reliability \& Recovery}} \\[1pt]
Reliability control packets (SACK, NACK) & Selective and negative per-packet acknowledgment mechanisms  & M \\
Trimmed packet support               & In-network packet truncation to headers and forwarding to the destination  & O \\
Reliability probes                   & Requester-initiated probes to query the responder's packet reception state       & O \\
Per-packet timer                     & Independent retransmit timer per outstanding packet    & O \\
Linear + Exponential ACK timeout     & Retransmit timer scales linearly before transitioning to exponential backoff & M \\
Fast-loss detection hook             & Implementation-specific fast loss detection and recovery   & O \\
Differentiated DSCP traffic classes  & Dedicated DSCP codepoints to isolate data, retransmission, and control packets    & M \\[3pt]

\multicolumn{3}{l}{\bf \textit{Congestion Control \& Load Balancing}} \\[1pt]
Network Signaled CC (NSCC)              & Window-based, SACK-clocked ECN+RTT algorithm over best-effort Ethernet    & O \\
Timestamp header                     & Requester-injected timestamps reflected by the responder for RTT measurement     & O \\
Service-time reporting/compensation  & Measures and subtracts responder host processing overhead from end-to-end RTT     & O \\
Responder host backpressure          & Propagates responder host-side congestion back to the requester         & O \\[3pt]

\multicolumn{3}{l}{\bf \textit{Resilience \& Fast Fail-over}} \\[1pt]
EV Probes                            & Endpoint-scope request/response exchange for path-scope liveness verification   & O \\
Port Status Update                   & Asynchronous endpoint signaling exposing local link-state updates               & O \\[4pt]
\bottomrule
\end{tabularx}
\vspace{-0.07in}
\end{table*}

\section{Design Overview}
\label{sec:design}
MRC modifies and extends RC to support per-connection multipath packet
spraying and fast failover under link and fabric faults while
preserving RC's reliable, ordered delivery and completion software
abstraction.  MRC combines concepts and work developed in
UltraEthernet~\cite{uet-overview, uet} and IBTA~\cite{ibta18} for this goal. The design is
organized as a set of orthogonal primitives that compose into a
coherent control loop: endpoints select per-packet paths, bound outstanding
traffic to control reordering and buffering, exchange selective
reliability and congestion feedback, and mitigate port and path
failures by broadcasting datapath-visible reachability signals.

Table~\ref{tab:mechanisms} categorizes MRC primitives; mandatory are
required for correctness while optional improve efficiency and
resilience.  Implementations adapt optional primitives depending on
hardware design point and expected network configuration.

These primitives are tightly co-designed: an Entropy Value (EV)
dictates which path a packet takes; EV spraying balances load across
all paths but introduces reordering; Maximum PSN Range (MPR) bounds
reordering and retransmit state; SACK/NACK/Trimmed packet feedback
keeps recovery tractable; reliability probes elicit connection state
and SACK-clocked congestion control uses EV reflection and ECN to
fine-tune load balancing and avoid failed paths. Finally, traffic
classes isolate control traffic from congestion and prioritize
retransmits, while EV probes and Port Status Updates signal path and
port health at line speed.

MRC deliberately simplifies the transport to support AI training. It
narrows data-plane operations to Write and Write-with-Immediate
(WriteImm) and removes RC end-to-end flow control in favor of explicit
bounded-flight mechanisms. MRC permits out-of-order data placement at
the responder (destination) to tolerate packet spraying and decouples
packet delivery from semantic processing; requester and responder
completion semantics remain unchanged.

\subsection{Multipath and Multiplane Operation}
\label{sec:multipath}

RC assigns each QP to one network path by ECMP hashing over the
connection 5-tuple. This leaves multipath fabric capacity
underutilized and provides no way for the endpoint to steer,
rebalance, or respond to path-level congestion or failure.

In contrast, MRC provides explicit, per-connection multipath through packet
spraying: the requester includes an Entropy Value (EV) field in each
packet and varies it on a per-packet basis so that packets from a
single QP are distributed across many fabric paths. EVs are a
first-class transport primitive: carried on the wire and interpreted
by the network to select a path and managed by endpoints for the
purposes of load balancing and resilience.

MRC supports three EV mechanisms: legacy ECMP hashing as used today,
Structured EV that encodes a deterministic hop-by-hop source routing
within the UDP source port and IPv6 flow label and SRv6 uSID source
routing~\cite{rfc8986, rfc9800} which supports explicit path selection
using SRv6 micro-segments.  All three share a common EV abstraction
allowing systems to seamlessly swap between implicit hashing and
explicit source routing without altering transport-layer multipath
logic, contingent solely on NIC capabilities.

EV selection is configured through EV profiles, which define the EV
“universe” available to a connection and how EVs are generated
(explicit lists, NIC-generated EVs, or SRv6-derived EVs). Profiles are
controller-managed and can be shared across multiple QPs, bounding
per-QP state. At runtime, implementations maintain lightweight per-EV
state to bias selection away from congested or unhealthy paths. Each
EV can be classified into one of \texttt{\small GOOD}, \texttt{\small
  SKIP}, \texttt{\small DENIED}, or \texttt{\small ASSUMED\_BAD}
states. Only EVs in the \texttt{\small GOOD} state are used for
transmission; the remaining states represent progressively stronger
exclusion, triggered by datapath observations or by explicit
controller intervention.  Implementations may also support EVs
transitioning back from \texttt{\small ASSUMED\_BAD} by EV probing
(\S\ref{sec:resilience}).  In Clos-style networks, this EV-based active
end-to-end load balancing is able to rapidly avoid congested or flaky
links, even beyond the point on the path where switches no longer
have any alternative paths to the destination.

MRC extends packet spraying to multiple fabric planes via multiport
NICs, partitioning EVs across physical ports (planes). This allows a
single QP to spray traffic across multiple physical fabric planes,
maximizing path diversity and isolating failures to a single port or
plane. Additionally, MRC decouples response path selection from the
forward request path. This routing asymmetry allows control packets to
utilize different paths and planes from data, reinforcing resilience
against partial fabric failures and asymmetric congestion.  The aim is
that network failures should not cause QP failure, or even greatly
impact the tail latency of transfers, as the performance of large
synchronous AI training jobs is strongly dictated by $100^{th}$
percentile transfer performance.

\subsection{Receiver Bounded Transmission}
\label{sec:inflight}
Sprayed packets often arrive out-of-order at the responder. Without
structural bounds on outstanding traffic, endpoints face responder
buffer exhaustion and requestor retransmission state explosion. MRC
mitigates this via the \emph{Maximum PSN Range} (MPR) -- a sliding
receive window that strictly bounds inflight request packets: a
requestor cannot send a packet with a sequence number beyond the upper
edge of the responder's advertised packet tracker bitmap.  Negotiated
during connection establishment, MPR reflects the responder's request
packet buffering capacity.

Both endpoints maintain a PSN-fidelity MPR-sized bitmap window to
track inflight packets. This structure underpins MRC’s selective
acknowledgment and retransmission mechanisms
(\S\ref{sec:reliability}), bounding the requestor's retransmit state
directly to the size of the negotiated MPR. Similarly, responder
out-of-order reception state is proportional to MPR.

While MPR can remain static throughout a connection's lifetime, MRC
optionally supports \emph{Dynamic MPR}, allowing the responder to
update the window size dynamically via SACKs. Dynamic MPR optimizes
hardware resource allocation along two axes: first, it elastically
scales bitmap memory based on traffic demand rather than worst-case
static provisioning; second, it allows the NIC to oversubscribe and
multiplex finite bitmap capacity across concurrent QPs by reclaiming
allocations from idle or low-rate connections.

In addition, an MRC responder advertises a limit on the number of concurrent
outstanding WriteIMM operations.  Whereas MPR bounds the
PSN window at request packet fidelity, WriteIMM limiting
bounds the number of semantic-level operations the responder is
prepared to process concurrently. Together, these decoupled mechanisms
provide independent dimensions of inflight bounding, enabling
responders to limit required packet and semantic operation buffering
and state.

\subsection{Reliability and Recovery}
\label{sec:reliability}

MRC targets best-effort Ethernet, where loss and reordering are
expected.  It introduces a packet delivery layer that tracks packet
reception at the responder decoupled from RDMA semantic processing.

\textbf{SACK and NACK}: MRC enforces packet reliability via Selective
Acknowledgments (SACK). Each SACK carries a cumulative acknowledgment,
a bitmap offset, and a bitmask of out-of-order arrivals relative to
that offset. This mechanism enables the requestor to explicitly
distinguish packet loss from transient reordering. The responder
advances the cumulative acknowledgment as gaps fill, sliding the
requestor's transmission horizon forward. Complementing this, Negative
Acknowledgments (NACKs) provide proactive non-delivery signaling to
trigger immediate retransmissions. Operating at the packet delivery
layer, NACKs are driven by deterministic events such as trimmed packet
arrival or local resource exhaustion.


To minimize head-of-line blocking and accelerate window advancement,
MRC optimizes for retransmission of the earliest missing
PSNs. Responders prioritize reporting the oldest incomplete regions of
their bitmap, while requesters retransmit using a differentiated,
higher-priority traffic class (\S\ref{sec:resilience}).

\textbf{Trimmed Packet Support:} MRC supports packet trimming where
switches truncate logically dropped packets, forwarding only their
headers via a high-priority traffic class. Responders process these
headers to generate a NACK to the requester. This explicit,
low-latency loss signal enables fast retransmissions that bypass
retransmit-timeout timers.

\textbf{Reliability Probes:} Requesters can issue out-of-band
reliability probes to query the responder's current connection
delivery state. Probes do not consume PSNs or alter connection state
because replies are decoupled from forward progress. They can be used
as a lightweight, fire-and-forget mechanism or via timers. Upon
receipt, responders return a standard SACK. Probes enable the
requestor to audit connection health proactively and independent of
relying on ACK timeouts.

\textbf{Timeout and Loss Detection:} MRC also refines the RC
connection timer and Local ACK timeout primitives, adding a
higher-resolution Local ACK timeout featuring a linear-to-exponential
backoff, alongside an optional per-packet timer
capability. Additionally, MRC allows implementation-specific loss
detection and recovery algorithms. Further details are available in
the specification~\cite{mrc-spec}.

\subsection{Congestion Control and Load Balancing}
\label{sec:cc}

MRC is designed to support Congestion Control (CC) over best-effort
Ethernet with focus on Network Signaled CC~\cite{uet}
(NSCC), a sender-based window-driven algorithm that
utilizes ECN and RTT-derived queueing delay to regulate a
byte-fidelity congestion window.  MRC standardizes the transport
signals required to drive NSCC.  Every SACK encapsulates CC metadata
within a dedicated sub-header that is consumed by the algorithm. This
state conveys forward-path ECN markings, responder accounting metrics
(e.g., cumulative bytes received) and responder-side congestion window
penalties. MRC also supports alternative algorithms that utilize a
subset of the standardized signals and metadata. Additionally, the
controller enables per-connection CC profile configuration, similar to
the EV profile.

RTT estimation is supported via two mechanisms: requesters can either
maintain local state or embed explicit timestamps within packet
headers for the responder to reflect back in the CC sub-header. MRC
also supports service-time compensation on a per-connection
basis. When enabled, the responder reports its internal processing
latency, allowing the requester to subtract this overhead and
accurately calculate network RTT even with responder processing
delays.

MRC supports host backpressure, enabling responders to signal
memory-subsystem queuing in an implementation-agnostic manner. NSCC
leverages this feedback to modulate the requester's congestion window,
preventing responder-side memory contention from degrading end-to-end
performance.

Finally, responders reflect forward-path EVs within return SACKs,
enabling the requester to correlate ECN markings with specific network
paths. Implementation-specific multipath and multiplane
load-balancing algorithms leverage this per-path feedback to
dynamically adjust path selection probabilities
(\S\ref{sec:multipath}).

\subsection{Resilience and Fast Fail-over}
\label{sec:resilience}

Large-scale network fabrics are prone to frequent disruptions such as
link flaps, blackholes, and routing asymmetries.  These degrade
performance and may even cause job crashes. Baseline RC is
ill-equipped to mitigate these issues, relying instead on routing
control-plane convergence and end-to-end timeouts.

MRC introduces \emph{endpoint operations}: lightweight, GID-scoped
control exchanges identified on the wire by a reserved QP identifier
(\texttt{0x2}). Receivers process these operations statelessly without
connection context, generating replies by simply swapping Layer 2 and
3 fields. GID scoping enables a single exchange to simultaneously
update state across all connections sharing that GID, amortizing
operation overhead and improving scalability.

MRC defines two endpoint operations: \textbf{EV probes} are
path-scoped and validate the viability of specific EVs.  They are used
primarily to drive EV selection and health
state~(\S\ref{sec:multipath}). Implementations may issue EV probes from
the datapath or via the controller API (\S\ref{sec:api})
depending on the deployment.

\textbf{Port status update} is an asynchronous operation that
announces port-level reachability in multiplane networks. It carries a
\texttt{\small port\_status\_mask} bitmap reflecting the sender's
local port health; the receiver logs this state to avoid degraded
ports. Port Status Updates may be issued directly from the datapath or
by software through the controller API.  This mechanism accelerates
reachability signaling from slow control-plane timescales to the
round-trip-time of the data plane.

Together, these mechanisms define a fast failover capability: Port
Status Updates proactively signal link and fabric failures, while EV
probes autonomously and continually query the viability of degraded
paths without control-plane intervention. By exposing both primitives
via the controller API as well as datapath, implementations can tailor
designs to the demands of the use-case and deployment.

\section{Wire Protocol}
\label{sec:wire}

MRC's wire protocol is a minimal extension over RoCEv2 RC: a new
opcode space, minor BTH modifications, new control packet types, and a
new class of endpoint operations, supported by new headers
(Table~\ref{tab:headers}) encapsulated within standard
RoCEv2/UDP/IP. MRC UDP destination port and checksum is consistent
with RoCEv2, and requires parsers to additionally inspect the DSCP
field.

\textbf{Opcode Space:} MRC isolates its packet types from RC by using
a distinct \texttt{0101} transport prefix. RC and MRC endpoints are
non-interoperable.

\textbf{BTH Modifications:} MRC modifies the BTH to add an
\texttt{\small rtx} bit flagging retransmissions, a \texttt{\small tsh} bit
signaling a TSETH header is present, and repurposes the PSN field as a
requester-local \texttt{\small request\_id} for probe/endpoint correlation.

\textbf{Request Packets:} Request packets share a uniform header stack
(Table~\ref{tab:headers}): BTH $\rightarrow$ METH $\rightarrow$
[TSETH] $\rightarrow$ RETH $\rightarrow$ [ImmDt]. RETH
is present in \emph{every} packet and METH is populated only
for immediate operations. Per-packet TSETH presence is signaled
via the BTH \texttt{\small tsh} bit; ImmDt is appended post-RETH
for immediate variants.

\textbf{Control Packets:} MRC bifurcates control traffic into
reliability control packets and endpoint operations, both of which
utilize new headers (Table~\ref{tab:headers}) appended to the BTH.
Reliability SACKs (SETH $\rightarrow$ CC\_STATE) convey delivery state
and congestion telemetry, NACKs (NETH) provide non-delivery signaling,
and Reliability Probes (PETH) elicit a SACK-encoded response. Endpoint
operations map Endpoint Requests (ERTH) directly to Endpoint Responses
(EETH).


\begin{table}[t]
\centering
\caption{MRC Header Modifications\vspace{-0.07in}}
\label{tab:headers}
\renewcommand{\arraystretch}{1.1}
\small
\begin{tabularx}{\columnwidth}{l l X}
\toprule
\textbf{Header} & \textbf{Status} & \textbf{Comment} \\
\midrule
BTH       & Modified      & Adds rtx, tsh bits; overloads PSN for probes and endpoint ops                 \\
TSETH     & New           & Timestamp / service-time                                                      \\
RETH      & Recast        & Adapted for MRC WRITE semantics                                               \\
METH      & New           & Message Header; tracks WRIMM ops                                              \\
SETH      & New           & SACK header; carries CC\_STATE                                                \\
NETH      & New           & NACK header                                                                   \\
PETH      & New           & Probe request header                                                          \\
ERTH      & New           & Endpoint request header                                                       \\
EETH      & New           & Endpoint response header                                                      \\
CC\_STATE & New           & Congestion control telemetry                                                  \\
ImmDt     & Unmod.        & Immediate Data                                                                \\
\bottomrule
\end{tabularx}
\end{table}

\section{Application and Controller API}
\label{sec:api}

MRC defines two APIs~\cite{mrc-github}: an application API
(\texttt{\small mrc.h}) and a privileged controller API
(\texttt{\small mrc\_ctl.h}).

\textbf{Application API:} Applications use \texttt{libibverbs} for
device discovery and memory registration, with MRC introducing new
entry points for transport control- and data-plane operations. MRC resources
(\texttt{\small mrc\_cq} and \texttt{\small mrc\_qp}) mirror verbs naming and
lifecycle conventions. Functions are restricted to transport-essential
operations while maintaining strict signature and parameter parity
with verbs. \texttt{\small mrc\_modify\_qp()} serves as the primary
configuration primitive, managing Dynamic MPR, Trimmed Packet support,
service-time compensation and EV/CC profile assignment. Extended
queries (\texttt{\small mrc\_query\_device()}, \texttt{\small mrc\_query\_port()})
expose MRC-specific capability flags and per-port features absent from
the verbs API. QP connection setup occurs out-of-band; every QP is
bound to EV and CC profiles pre-configured using the controller API.

\textbf{Controller API:} The controller runs as a privileged process
(\texttt{\small CAP\_NET\_ADMIN}) to govern policy and telemetry. Mandatory
capabilities are device/port queries and EV/CC profile management.
Optional features are EV state events and EV probes.

\section{Conclusion}
\label{sec:conclusion}
MRC demonstrates that the multipath, reliability, congestion control
and resilience deficiencies of RoCEv2 RC are resolvable via a
practical, incremental and composable extension. MRC builds on
existing hardware designs to deliver an open~\cite{mrc-spec},
production-ready transport for large-scale AI/ML training; a
comprehensive empirical evaluation is provided in a companion
paper~\cite{mrc-nsdi}.

\section*{Acknowledgments}
Large Language Models (LLMs) were utilized for language editing during
the preparation of this manuscript.

\balance
\bibliographystyle{IEEEtran}
\bibliography{biblio}

@misc{dhen,
      title={{DHEN}: A Deep and Hierarchical Ensemble Network for Large-Scale Click-Through Rate Prediction}, 
      author={Buyun Zhang and Liang Luo and Xi Liu and Jay Li and Zeliang Chen and Weilin Zhang and Xiaohan Wei and Yuchen Hao and Michael Tsang and Wenjun Wang and Yang Liu and Huayu Li and Yasmine Badr and Jongsoo Park and Jiyan Yang and Dheevatsa Mudigere and Ellie Wen},
      year={2022},
      eprint={2203.11014},
      archivePrefix={arXiv},
      primaryClass={cs.IR}
}

@inproceedings{IRN,
author = {Mittal, Radhika and Shpiner, Alexander and Panda, Aurojit and Zahavi, Eitan and Krishnamurthy, Arvind and Ratnasamy, Sylvia and Shenker, Scott},
title = {Revisiting Network Support for {RDMA}},
year = {2018},
isbn = {9781450355674},
publisher = {Association for Computing Machinery},
address = {New York, NY, USA},
url = {https://doi.org/10.1145/3230543.3230557},
doi = {10.1145/3230543.3230557},
booktitle = {Proceedings of the 2018 Conference of the ACM Special Interest Group on Data Communication},
pages = {313–326},
numpages = {14},
location = {Budapest, Hungary},
series = {SIGCOMM '18}
}

@article{dnntraining-overview,
author = {Ben-Nun, Tal and Hoefler, Torsten},
title = {Demystifying Parallel and Distributed Deep Learning: An In-depth Concurrency Analysis},
year = {2019},
issue_date = {July 2020},
publisher = {Association for Computing Machinery},
address = {New York, NY, USA},
volume = {52},
number = {4},
issn = {0360-0300},
url = {https://doi.org/10.1145/3320060},
doi = {10.1145/3320060},
journal = {ACM Comput. Surv.},
month = {aug},
articleno = {65},
numpages = {43}
}

@inproceedings{falcon,
author = {Singhvi, Arjun and Dukkipati, Nandita and Chandra, Prashant and Wassel, Hassan M. G. and Sharma, Naveen Kr. and Rebello, Anthony and Schuh, Henry and Kumar, Praveen and Montazeri, Behnam and Bansod, Neelesh and Thomas, Sarin and Cho, Inho and Seibert, Hyojeong Lee and Wu, Baijun and Yang, Rui and Li, Yuliang and Huang, Kai and Yin, Qianwen and Agarwal, Abhishek and Vaduvatha, Srinivas and Wang, Weihuang and Moshref, Masoud and Ji, Tao and Wetherall, David and Vahdat, Amin},
title = {Falcon: A Reliable, Low Latency Hardware Transport},
year = {2025},
isbn = {9798400715242},
publisher = {Association for Computing Machinery},
address = {New York, NY, USA},
url = {https://doi.org/10.1145/3718958.3754353},
doi = {10.1145/3718958.3754353},
booktitle = {Proceedings of the ACM SIGCOMM 2025 Conference},
pages = {248–263},
numpages = {16},
location = {S\~{a}o Francisco Convent, Coimbra, Portugal},
series = {SIGCOMM '25}
}

@misc{llama3,
      title={The Llama 3 Herd of Models}, 
      author={Aaron Grattafiori et al},
      year={2024},
      eprint={2407.21783},
      archivePrefix={arXiv},
      primaryClass={cs.AI},
      url={https://arxiv.org/abs/2407.21783}, 
}

@inproceedings{meta-roce,
author = {Gangidi, Adithya and Miao, Rui and Zheng, Shengbao and Bondu, Sai Jayesh and Goes, Guilherme and Morsy, Hany and Puri, Rohit and Riftadi, Mohammad and Shetty, Ashmitha Jeevaraj and Yang, Jingyi and Zhang, Shuqiang and Fernandez, Mikel Jimenez and Gandham, Shashidhar and Zeng, Hongyi},
title = {RDMA over Ethernet for Distributed Training at Meta Scale},
year = {2024},
isbn = {9798400706141},
publisher = {Association for Computing Machinery},
address = {New York, NY, USA},
url = {https://doi.org/10.1145/3651890.3672233},
doi = {10.1145/3651890.3672233},
booktitle = {Proceedings of the ACM SIGCOMM 2024 Conference},
pages = {57–70},
numpages = {14},
location = {Sydney, NSW, Australia},
series = {ACM SIGCOMM '24}
}

@misc{uet-overview,
      title={Ultra Ethernet's Design Principles and Architectural Innovations}, 
      author={Torsten Hoefler and Karen Schramm and Eric Spada and Keith Underwood and Cedell Alexander and Bob Alverson and Paul Bottorff and Adrian Caulfield and Mark Handley and Cathy Huang and Costin Raiciu and Abdul Kabbani and Eugene Opsasnick and Rong Pan and Adee Ran and Rip Sohan},
      year={2025},
      eprint={2508.08906},
      archivePrefix={arXiv},
      primaryClass={cs.NI},
      url={https://arxiv.org/abs/2508.08906}, 
}

@misc{uet,
      title={Ultra Ethernet Specification v1.0.1},
      author={{Ultra Ethernet Consortium}},
      year={2025},
      url={https://ultraethernet.org/wp-content/uploads/sites/20/2025/10/UE-Specification-1.0.1.pdf}
}

@misc{rfc8986,
  series = {Request for Comments},
  number = {8986},
  howpublished = {RFC 8986},
  publisher = {RFC Editor},
  doi = {10.17487/RFC8986},
  url = {https://www.rfc-editor.org/info/rfc8986},
  author = {Clarence Filsfils and Darren Dukes and Stefano Previdi and John Leddy and Satoru Matsushima and Daniel Voyer},
  title = {{Segment Routing over IPv6 (SRv6) Network Programming}},
  year = {2021},
  month = feb,
}

@misc{rfc9800,
  series = {Request for Comments},
  number = {9800},
  howpublished = {RFC 9800},
  publisher = {RFC Editor},
  doi = {10.17487/RFC9800},
  url = {https://www.rfc-editor.org/info/rfc9800},
  author = {Weiqiang Cheng and Clarence Filsfils and Zhenbin Li and Bruno Decraene and Dezhong Cai and Daniel Voyer and Francois Clad and Shay Zadok and Jim Guichard and Aihua Liu and Robert Raszuk and Cheng Li},
  title = {{Compressed SRv6 Segment List Encoding}},
  year = {2025},
  month = jun,
}

@manual{mrc-spec,
  title        = {{Open Compute Project: Multipath Reliable Connection (MRC) Specification, Version 1.0}},
  author       = {Rip Sohan and Eric Spada and Eric Davis and Mark Handle and Idan Burstein and Tony Hurson and Jithin Jose and Vivek Kashyap and Rong Pan and Sayantan Sur},
  organization = {Open Compute Project Foundation},
  year         = {2026},
  url          = {https://www.opencompute.org/documents/ocp-mrc-1-0-pdf},
  note         = {Accessed: May 17, 2026}
}

@misc{mrc-nsdi,
      title={{Resilient AI Supercomputer Networking using MRC and SRv6}},
      author={Joao Araujo and Alex Chow and Mark Handley and Ryder Lewis and Christoph Paasch and Jitendra Padhye and Michael Papamichael and Greg Steinbrecher and Amin Tootoonchian and Lihua Yuan and S. Anantharamu and Abhishek Dosi and Mohit Garg and Mahdieh Ghazi and Torsten Hoefler and Deepal Jayasinghe and Jithin Jose and Abdul Kabbani and Guohan Lu and Yang Wang and K. Doddapaneni and Murali Garimella and Vipin Jain and Yanfang Le and H. Nagulapalli and S. Narayanan and Rong Pan and Rathina Sabesan and Raghava Sivaramu and Rip Sohan and Eric Davis and Dragos Dumitrescu and Mohan Kalkunte and Bhaswar Mitra and Guglielmo Morandin and Adrian Popa and Costin Raiciu and Eric Spada and John Spillane and Niranjan Vaidya and Aviv Barnea and Idan Burstein and Elazar Cohen and Yamin Friedman and Noam Katz and Masoud Moshref and Yuval Shpigelman and Shahaf Shuler and Shy Shyman and Sayantan Sur},
      year={2026},
      eprint={2605.04333},
      archivePrefix={arXiv},
      primaryClass={cs.NI},
      url={https://arxiv.org/abs/2605.04333},
}

@inproceedings{lu2025alibaba,

  author       = {Jie Lu and Jiaqi Gao and Fei Feng and Zhiqiang He and Menglei
                  Zheng and Kun Liu and Jun He and Binbin Liao and
                  Suwei Xu and Ke Sun and Yongjia Mo and Qinghua Peng
                  and Jilie Luo and Qingxu Li and Gang Lu and Zishu
                  Wang and Jianbo Dong and Kunling He and Sheng Cheng
                  and Jiamin Cao and Hairong Jiao and Pengcheng Zhang
                  and Shu Ma and Lingjun Zhu and Chao Shi and Yangming
                  Zhang and Yiquan Chen and Wei Wang and Shuhong Zhu
                  and Xingru Li and Qiang Wang and Jiang Liu and Chao
                  Wang and Wei Lin and Ennan Zhai and Jiesheng Wu and
                  Qiang Liu and Binzhang Fu and Dennis Cai},
editor       = {Mar{\'{\i}}lia Curado and
                  Christian Esteve Rothenberg and
                  George Porter and
                  Srikanth Kandula},
  title        = {Alibaba Stellar: {A} New Generation {RDMA} Network for Cloud {AI}},
  booktitle    = {Proceedings of the {ACM} {SIGCOMM} 2025 Conference, {SIGCOMM} 2025,
                  S{\~{a}}o Francisco Convent, Coimbra, Portugal, September 8-11, 2025},
  pages        = {453--466},
  publisher    = {{ACM}},
  year         = {2025},
  url          = {https://doi.org/10.1145/3718958.3750539},
  doi          = {10.1145/3718958.3750539},
  bibsource    = {dblp computer science bibliography, https://dblp.org}
}

@manual{ibta18,
  title        = {InfiniBand Architecture Specification, Volume 1, Release 1.8},
  author       = {{InfiniBand Trade Association (IBTA)}},
  organization = {InfiniBand Trade Association},
  year         = {2025},
  url          = {https://www.infinibandta.org}
}

@misc{mrc-github,
  title        = {{OCP Multipath Reliable Connection}},
  author       = {{Open Compute Project}},
  year         = {2026},
  url          = {https://github.com/opencomputeproject/OCP-Multipath-Reliable-Connection},
  note         = {Accessed: May 18, 2026}
}

\end{document}